\documentclass[a4paper,conference]{IEEEtran}

\usepackage[cmex10]{amsmath}
\usepackage{amsthm}
\usepackage{cite}
\usepackage{amssymb}
\usepackage[caption=false]{subfig}
\usepackage{graphicx}
\usepackage{amsfonts}
\usepackage{cases}
\usepackage[top=122pt, bottom=54pt, left=54pt, right=37pt]{geometry}

\begin{document}

\title{Robust Precision Positioning Control on Linear Ultrasonic Motor}


\author{\authorblockN{Minh H-T Nguyen\authorrefmark{1}\authorrefmark{2}, Kok Kiong Tan\authorrefmark{1}\authorrefmark{2}, Wenyu Liang\authorrefmark{1}\authorrefmark{2}, Chek Sing Teo\authorrefmark{1}\authorrefmark{3}} 
\authorblockA{\authorrefmark{1}SIMTech-NUS Joint Laboratory (Precision Motion Systems), National University of Singapore, Singapore, 117576}
\authorblockA{\authorrefmark{2}Department of Electrical and Computer Engineering, National University of Singapore, Singapore, 117576 }
\authorblockA{\authorrefmark{3}Singapore Institute of Manufacturing Technology, A*STAR, Singapore, 638075}
}

\maketitle

\begin{abstract}
Ultrasonic motors used in high-precision mechatronics are characterized by strong frictional effects, which are among the main problems in precision motion control. The traditional methods apply model-based nonlinear feedforward to compensate the friction, thus requiring closed-loop stability and safety constraint considerations. Implementation of these methods requires computation power. This paper introduces a systematic approach using piecewise affine models to emulate the friction effect of the motor motion. The well-known model predictive control method is employed to deal with piecewise affine models. The increased complexity of the model offers a higher tracking precision on a simpler gain scheduling scheme.
\end{abstract}


\section{Introduction}

The ultrasonic motor (USM) is a type of piezoelectric actuator which uses some form of piezoelectric material and is implemented on the basis of piezoelectric effect. The USM offers advantages of high resolution and speed to ensure the precision and repeatability, so it is widely used in precision engineering, robots and medical/surgical instruments where high accuracy is required. Different from the typical piezoelectric actuator (PA) driven directly by the deformation of the piezoelectric material when a voltage applies, the USM provides motions by the friction between the piezoelectric material on the stator and the rotor. Thus, the USM offers another advantage of theoretically unlimited travel distance in comparison with the typical PA.

In the friction-based motion of USM, frictional forces are the main disturbance that degrades the closed loop performance. Because the friction presents a nonlinear switch which is dependent on the motion direction, using a single linear model to design a linear controller results in inaccuracy especially at low-speed control \cite{Jam09Friction}. Additionally, a practical controller should respect the physical limitations of the motor input and safety constraints on the system variables (e.g., position range, speed).

Most of the existing control designs to deal with the friction are based on decoupling the friction model from the linear motion system and mitigating it through a nonlinear model-based input beside a linear regulator such as PID. Along this perspective, much research efforts have been focusing on building accurate friction models \cite{Par04Identification,Pen05Modeling,Hay09Discrete}. The compensation, usually of bang-bang type in practice, resolves the friction problem and leaves PID with other unmeasured disturbances including the friction model mismatch. The approaches are simple to implement and if properly tuned, they provide fast transient response, good static accuracy and robustness to the motor parameter variations \cite{Pen07Intelligent}. However, the nonlinear compensation is contingent on asymptotic stability, which relies on the specified friction model. The frictional effects can also depend on rotor position and system degeneration, so a fixed friction model may require computing time to be evaluated. Finally, such control strategies do not systematically deal with constraints on the control input and variables, so manual safety designs must be implemented. .

A recent rising approach to deal with friction in electrical drive is based on piecewise affine (PWA) modeling of the nonlinear frictional effects. In \cite{Vas07Hybrid,Her09Minimum}, the authors modeled motion with static friction and applied it on model predictive control (MPC) to design time-optimal control strategies. The tracking performance is promising although the method still depends on the choice of friction models and no robustness is considered. Secondly, mixed-integer linear programming results in too many regions in the look-up tables \cite{Vas07Hybrid}, thus solution complexity is increased. 

This paper builds a robust optimal controller for ultrasonic motors. The commonly used friction models are approximated over linear segments directly by experiment data, thus nonlinear model identification in existing solutions is avoided. This model also describes Stribeck effect instead of simple static friction in \cite{Her09Minimum}. Specially, an integral MPC design imposes the robustness on model-plant mismatch near zero-speed to further mitigate the friction. Using quadratic programming simplifies the number of regions to look up for the system state. Finally, real-time control is implemented by a gain-scheduling table so the implementation complexity is comparable to the traditional feedforward PID.

In Section II, the ultrasonic drive and its discrete time hybrid model are described. We show how the piecewise affine model can fit in the description of the friction. Integral MPC with robustness design is presented in Section III. Simulation studies are described in Section IV before final implementation on the experiment setup is reported in Section V.

\subsection*{Notations}
$y,v$ denote position and velocity of the motor. $F,f$ are the general friction and its components. $A,B,C,D$ are matrices 
\newgeometry{top=104pt, bottom=54pt, left=54pt, right=37pt}
of a state space dynamics. Indices $i,j$ are the linear dynamics and subspace index, respectively. All sets mentioned in this context are polyhedral sets. 

\section{Ultrasonic Motor}

\subsection{System Description}
\begin{figure}[ht]%
\centering
\includegraphics[width=3.5in]{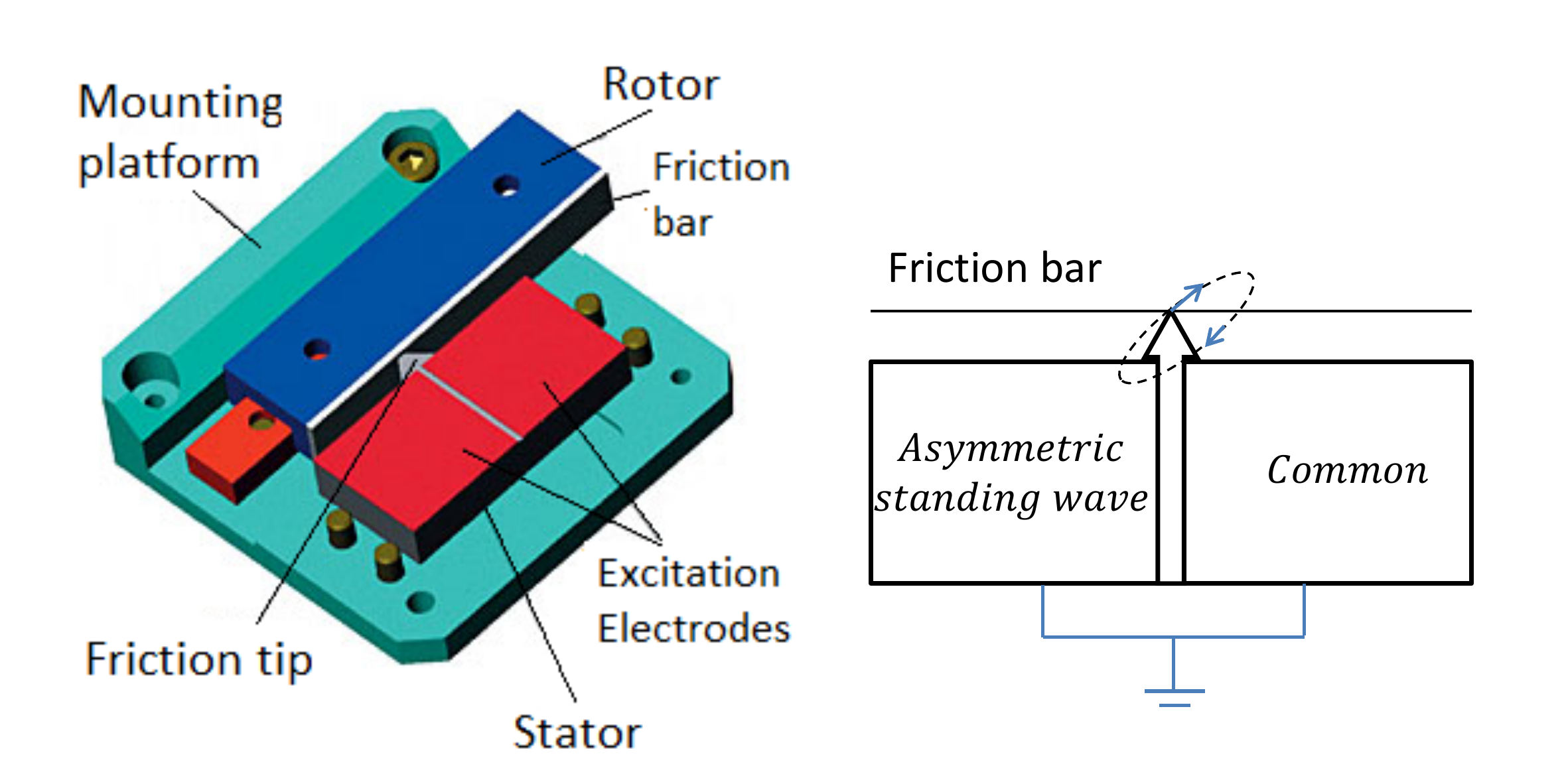}%
\caption{Linear ultrasonic motor structure and motion description.}%
\label{fig1a}%
\end{figure}

In this paper, the USM manufactured by \emph{Physik Instrumente (PI) GmbH \& Co. KG.} (model number M-663) is used. Its internal structure (without the encoder) and working principle are shown in Fig. \ref{fig1a}. Its motion is based on a alumina tip attached to the piezo-ceramic plate (the stator), segmented on one side by two electrodes. Depending on the desired direction of motion, the left or right electrode of the piezo-ceramic plate is excited with a standing wave to produce high-frequency vibration. Because of the asymmetric characteristic of the standing wave, the tip moves along an inclined linear path with respect to the friction bar surface and drives the rotor forward or backward. Each oscillatory cycle of the tip can transfer a $0.3\ \mu$m linear movement to the friction bar.  With the high-frequency oscillation, it will result in a smooth and continuous rotor motion. Besides, the drive C-185, manufactured by the same company, is used to convert analog input signals into the required high-frequency drive signals.

For control purpose, in the next section, the relationship between the input and the rotor motion is described by a normal friction-motion model, subject to certain operating constraints: travel range ($\pm9.5$ mm), velocity ($400$ mm/s) and input voltage ($\pm10\ V$).

\subsection{Piecewise Affine Model of Motion}
\begin{figure}[t]%
\centering
\includegraphics[width=3.5in]{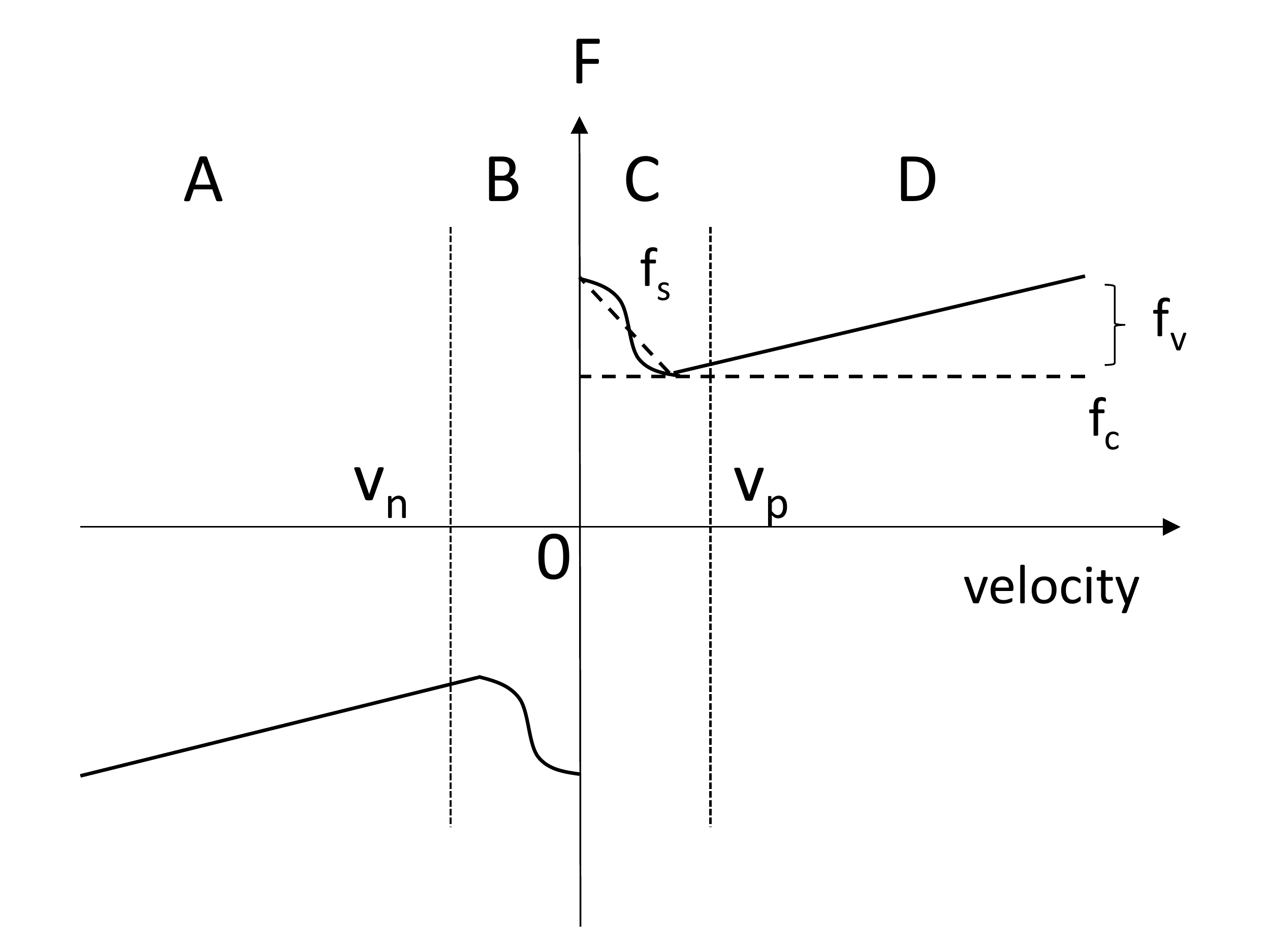}%
\caption{Motion friction described by linear segments over four regions from A to D.}%
\label{fig1}%
\end{figure}

\begin{figure}[t]%
\centering
\includegraphics[width=3.7in]{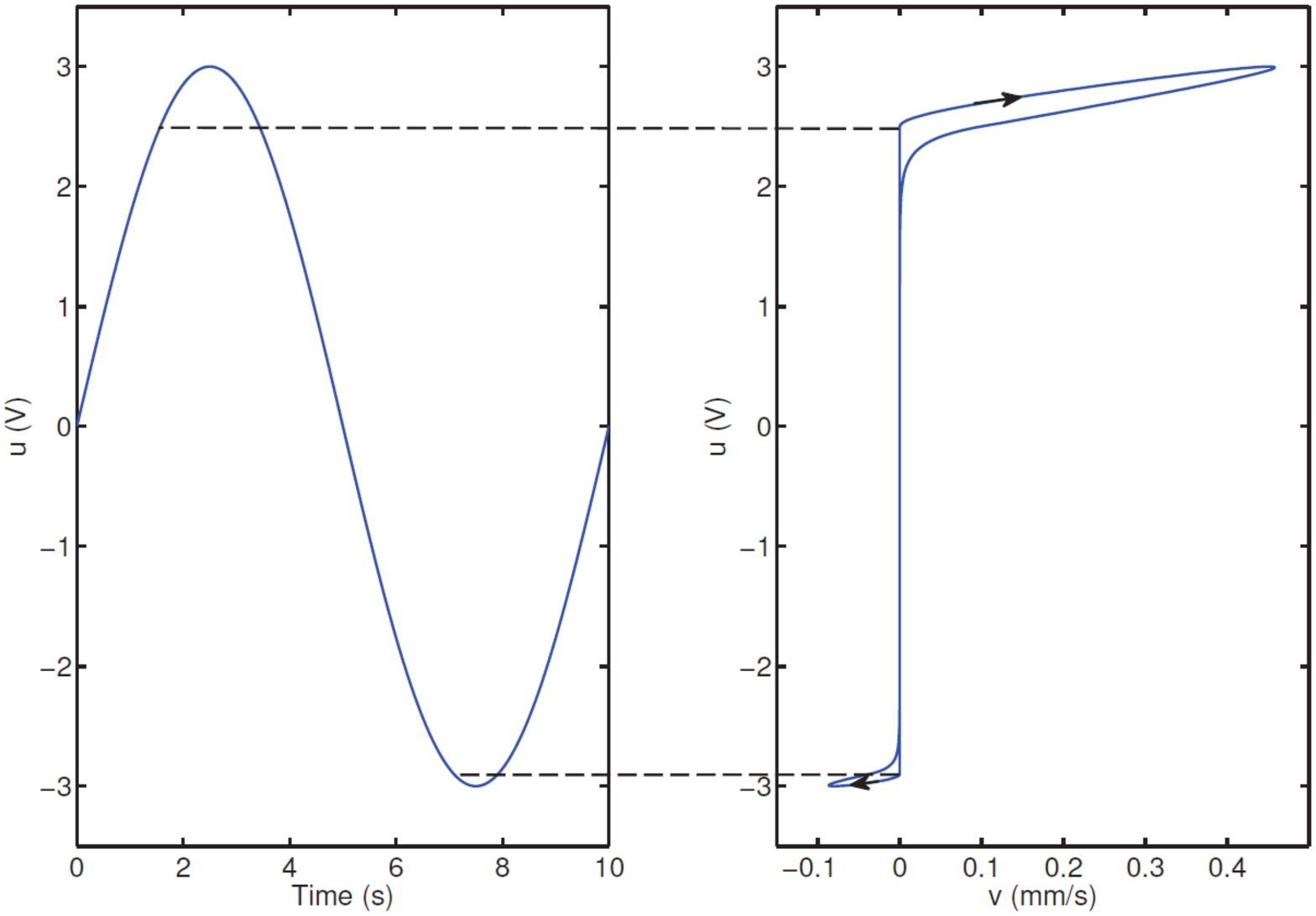}%
\caption{Identify static friction values using sine wave input.}%
\label{figfric}%
\end{figure}

\begin{figure*}[t]
\centering
\captionsetup[subfigure]{labelformat=empty}
\subfloat{\includegraphics[width=3.2in]{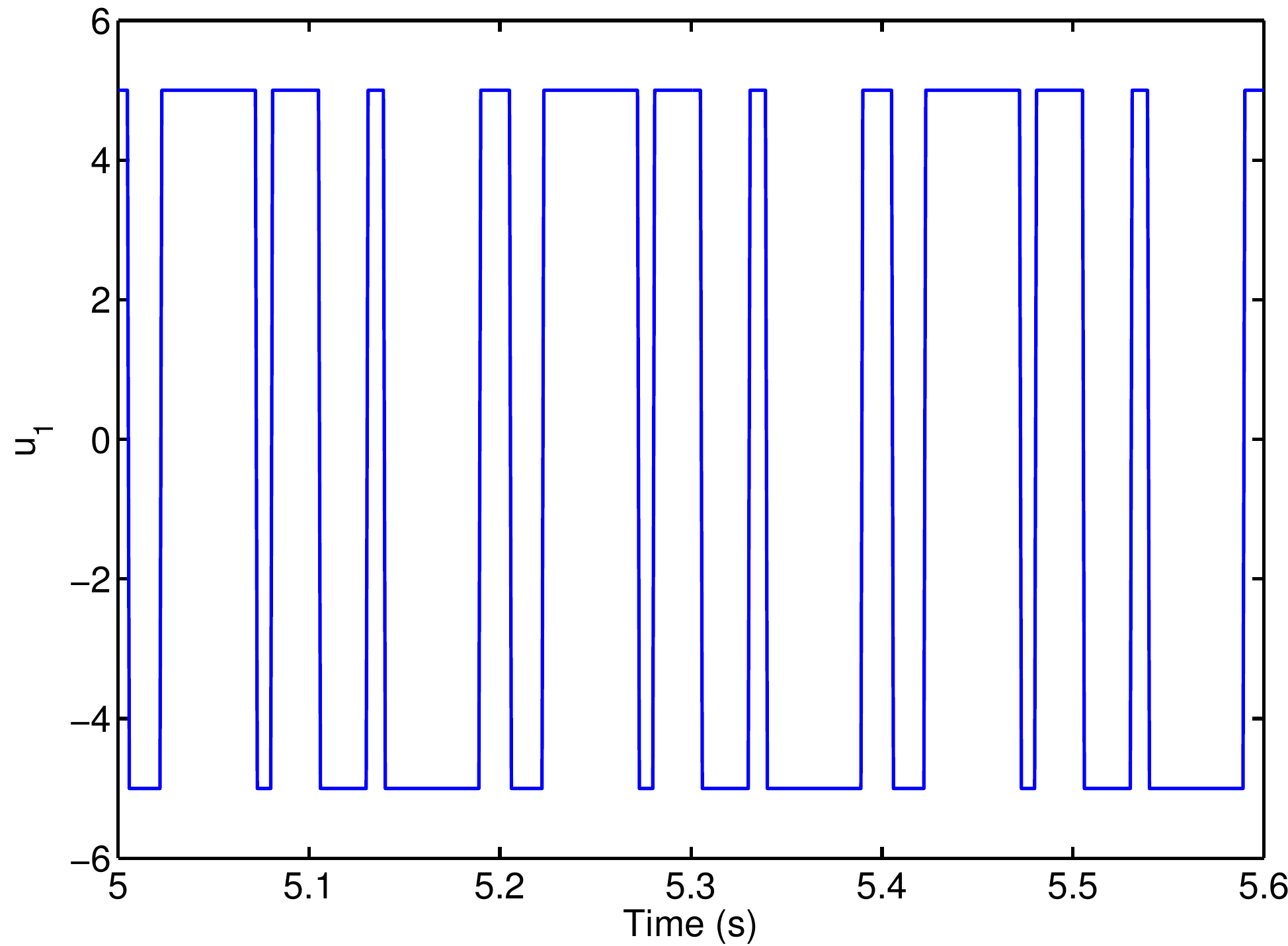}}
\subfloat{\includegraphics[width=3.2in]{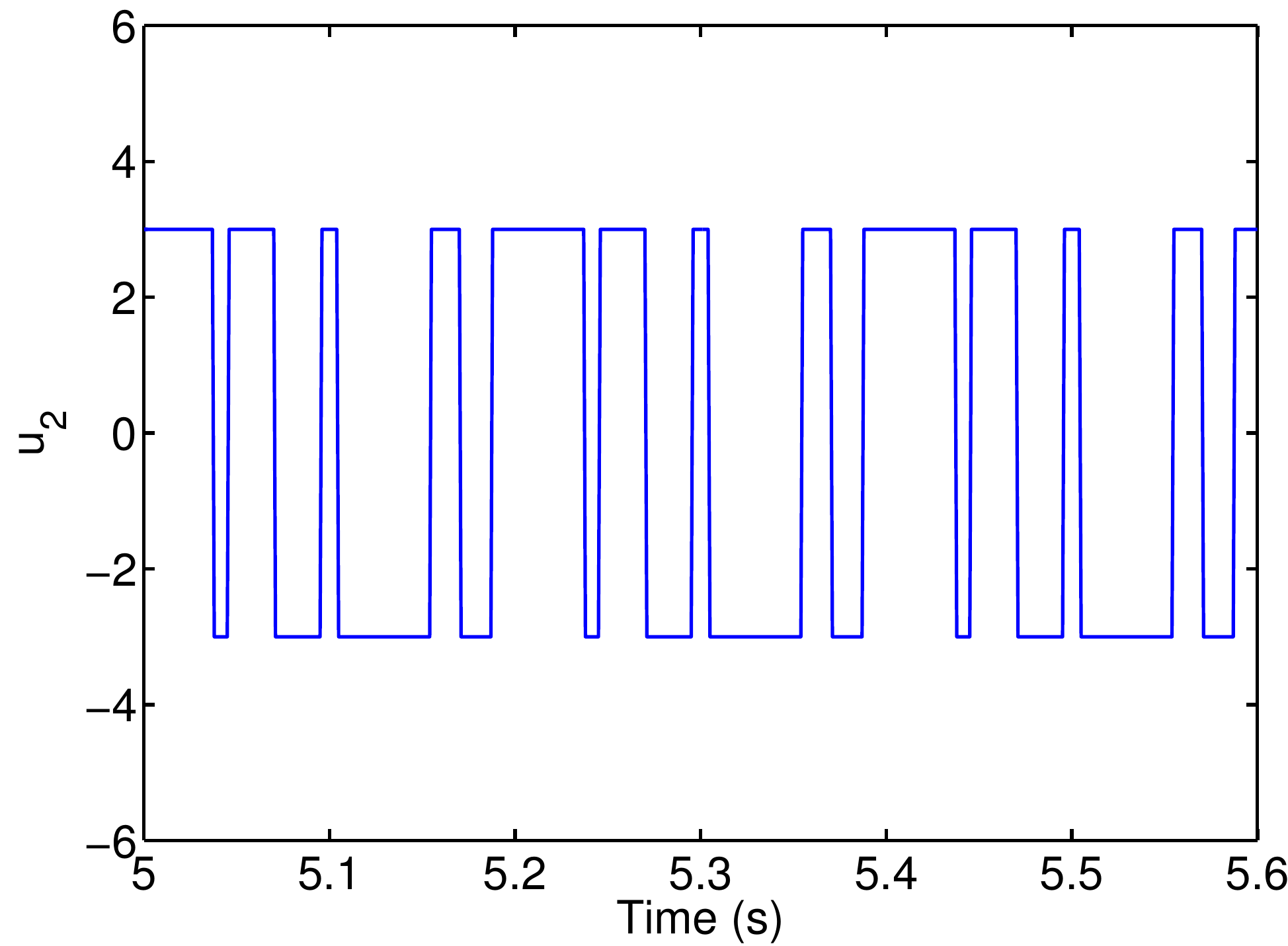}}\\
\subfloat[(a)]{\includegraphics[width=3.2in]{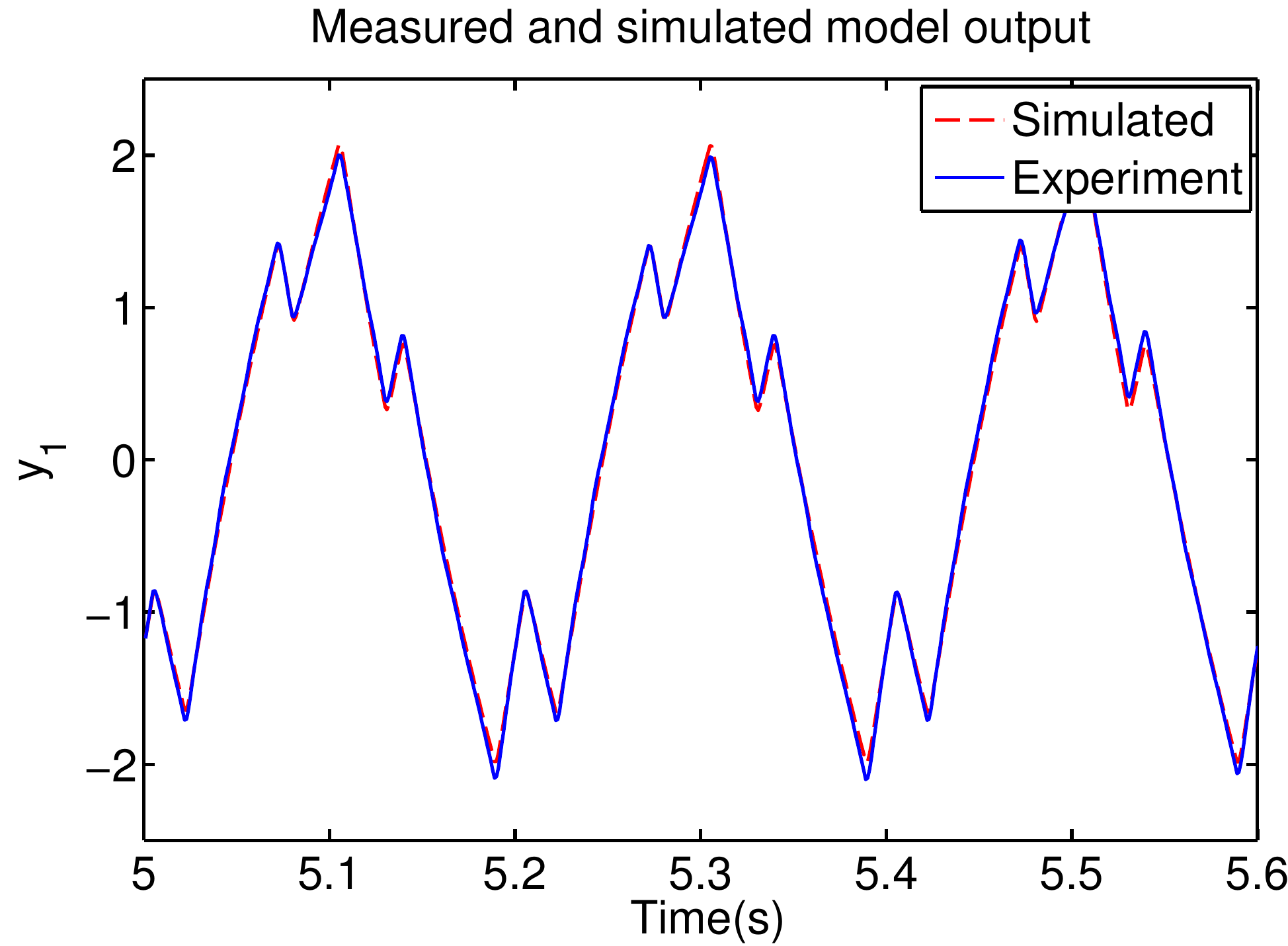}}
\subfloat[(b)]{\includegraphics[width=3.2in]{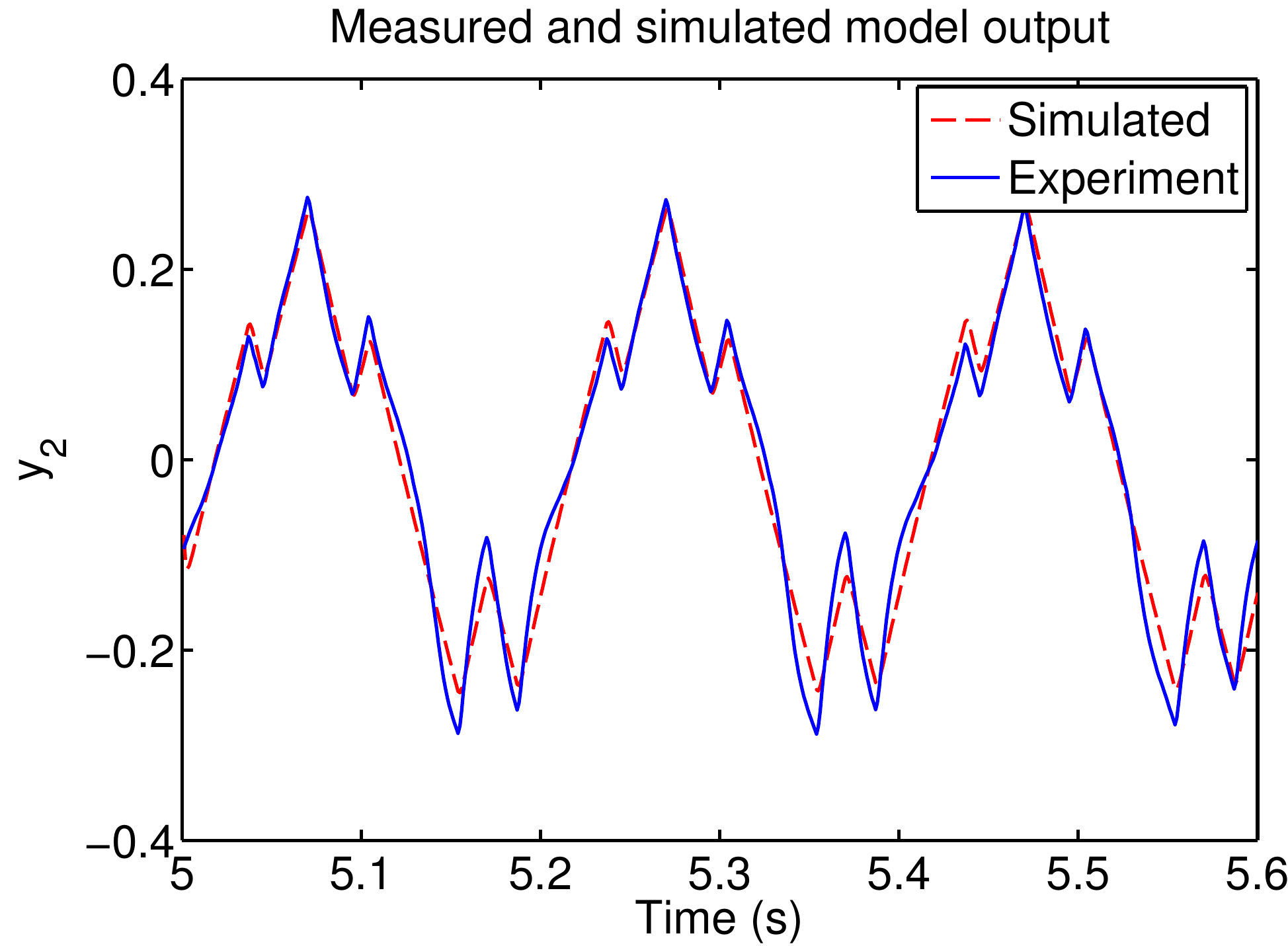}}
\captionsetup[subfigure]{labelformat=parens}
\caption{Model validation at two input ranges around $u_1 = 5$ (left) and $u_2 = 3$ (right)}
\label{fig2}
\end{figure*}

Consider a classical linear motion model which takes the form 
\begin{equation}
\begin{bmatrix}\dot{y}\\ \dot{v}\end{bmatrix} = \begin{bmatrix}0& 1\\ a& b\end{bmatrix}\begin{bmatrix}y\\ v\end{bmatrix}+\begin{bmatrix}0\\ c\end{bmatrix}(u-F(v))
\label{eq:linmdl}
\end{equation}
where $y,v$ is the rotor position and velocity; $F(v)$, the friction as shown in Fig. \ref{fig1}, consists of the constant Coulomb friction $f_c$, viscous friction $f_v = kv$ and Stribek effect $f_s$ showing how the friction continuously decreases as the motor accelerates. 

From there, the motion system of ultrasonic motor can be represented in four regions A-D, as in Fig. \ref{fig1}. Because the viscosity is linear in $v$, models in regions A and D can be represented by \eqref{eq:linmdl}. In regions B and C, the same structure can be employed to approximate the system, but with different linear dynamics; the complexity in the pre-sliding regime will be addressed by the robust design in Section III.

First, the relationship from input to rotor position are identified. This is done by modeling the function between effective inputs (input minus static friction) and the position output. Also, by assuming symmetry in the friction model, two input signals are required: one to model the motion in regions B and C and another to model the motion in regions A and D.  The correlation at very low speed and normal speed are measured. The asymmetric static friction values at which the motor starts moving, determined by injecting a sine wave function with low frequency and amplitude $3\ V$ , are $f_{cp}= 2.5\ V, f_{cn}= -2.9\ V$ as seen from Fig. \ref{figfric}.

Since the average static friction is $2.5 V$ and $-2.9 V$ respectively for positive and negative directions, test inputs with double frequency square waves and magnitude $u = \pm 5 V$ and $u = \pm 3 V$ are used to model the input-position correlation. By removing the friction force from the test inputs, the two models $\{A_i,B_i\},\ i=1,2)$ are obtained and validated in Fig. \ref{fig2}. The defining planes between the regions are taken at the velocity $v_n, v_p$ obtained by applying $u = \pm 3 \ V$ so the regions B and C safely encompass the nonlinear friction model. By removing the friction force from the test inputs, the two models $\{A_i,B_i\},\ i=1,2$ are obtained and validated in Fig. \ref{fig2}.

Denote the state $x=\begin{bmatrix}y & v\end{bmatrix}^T$. The piecewise affine model, after being transformed to discrete time, can be formally defined in the four convex subspaces
\begin{numcases}{x_{k+1} = }
A_1x_k+B_1(u_k-f_{cp}) & if $v\geq v_p \ (\Omega_1)$\nonumber\\
A_1x_k+B_1(u_k-f_{cn}) & if $v\leq v_n \ (\Omega_2)$\nonumber\\
A_2x_k+B_2(u_k-f_{cp}) & if $0\leq v\leq v_p \ (\Omega_3)$\IEEEeqnarraynumspace\\
A_2x_k+B_2(u_k-f_{cn}) & if $v_n\leq v\leq 0 \ (\Omega_4)$\nonumber
\label{pwasys}
\end{numcases}

\section{Integral MPC control}
In this section, the integral model predictive design is presented, consisting of state augmentation, MPC tracking formulation and MPC design.

Since this paper emphasizes on the position tracking, a new augmented system $(\bar{x}_k,\bar{u}_k,\bar{y}_k)$ is
\begin{IEEEeqnarray}{rCl}
\begin{bmatrix}x_{k+1}\\ r_{k+1}\\ \theta_{k+1}\end{bmatrix}&=&
\begin{bmatrix}A_i& B_i& 0\\ \textbf{0}& 1& 0\\C_i& -1& 1\end{bmatrix}
\begin{bmatrix}x_{k}\\ r_{k}\\ \theta_{k}\end{bmatrix}
+\begin{bmatrix}B_i\\ 1\\ 0\end{bmatrix}u_k\nonumber\\
\bar{y}_k&=&\begin{bmatrix}C_i& 0& 0\end{bmatrix}\bar{x}_k
\label{eq:augsys}
\end{IEEEeqnarray}
Due to the motion system characteristics that the model description already contains an integrator, it is not necessary to use $\Delta u$-tracking formula here. Instead, integrating state $\theta_k$ is introduced for zero-offset warranty. 

An MPC optimal control scheme uses the system \eqref{eq:augsys} to predict the output error ahead in time and uses current feedback errors to compensate any disturbance. The general form of MPC is stated as follows
\begin{IEEEeqnarray}{lCl}
V_N^o(\bar{x}_0,\tilde{U}&&)=\underset{\bar{U}}{\operatorname{min.}}\,\bar{x}_N^TP_j\bar{x}_N\nonumber\\
&&\qquad+\sum_{k=0}^{N-1}({\bar{x}_k}^T\bar{C}_j^TQ\bar{C}_j\bar{x}_k+\bar{u}_k^TR\bar{u}_k)\label{MPClaw}\\
subj.\,to \, && \bar{x}_k\in \mathbb{X}, \bar{u}_k\in \mathbb{U} \quad k=0,...,N-1;\ \bar{x}_N\in X_{if},\nonumber\\
&&\bar{u}_N=K_{j}\bar{x}_k+d_j \quad \quad \quad k\geq N\nonumber
\end{IEEEeqnarray}
where different prediction models in \eqref{pwasys} are used if $\bar{x}_k\in \Omega_j\ (j=1,2,3,4)$. $\mathbb{X}, \mathbb{U}$ are the state and input constraint set. After $N$ control steps, the scheme expects the state to reside inside the terminal regions $X_{jf}$, which is also an control invariant set defined by a linear state feedback $\bar{u}=K_{j}\bar{x}+d_j$ (with auxiliary input $d_j$). 

Stability analysis for hybrid (PWA) systems using MPC has been analyzed carefully in \cite{Laz06Stabilizing} where the design of terminal cost $P_j$ and terminal set $X_{jf}$ is proposed. In this paper, however, we more focus on the robust design aspect. Because regions A, D and B, C shares the same linear dynamics, the MPC design only needs to consider two cases for $\{A_1,B_1\}$ and $\{A_2,B_2\}$. The purpose is to keep the system state to stay within a terminal constraint invariant set inside the dynamic $\{A_2,B_2\}$ (near the switching surface of the friction) by a robust gain $K_2$. Such a gain can be designed specifically for the dynamic $P_2$ with bounded disturbance assumption. The rest of this section describes a unique MPC component design.

\emph{Terminal gain}:
Consider again the augmented dynamics from \eqref{pwasys}. Let $d_i=f_{cp}$ (or $f_{cn}$), take the feedback input $\bar{u}^{fb}=K_{i}\bar{x}$ as $K_1=K_{LQR}(Q,R)$ and $K_2$ such that the following system
\begin{IEEEeqnarray}{rCl}
\bar{x}_{k+1}&=&\bar{A}_2\bar{x}_k+\bar{B}_2(\bar{u}^{fb}_k+w_k)\nonumber\\
\bar{y}_{k}&=&\bar{C}_2\bar{x}_k
\end{IEEEeqnarray}
is robust against the friction model mismatch $w\ (|w|\leq w^*)$. This design can use one of many existing techniques in the literature to deal with input disturbance. For this application, a non-recursive method for $H_\infty$ \cite{Gad09H-Infinity} is applied to solve the related discrete-time Riccati equation (DARE). The obtained gain $K_2$ guarantees 
\begin{equation}
\left\|T_{wy}\right\|_{\infty}\leq\gamma
\label{robdes}
\end{equation}
where $T_{wy}$ is the transfer function from $w_k$ to $y_k$ and $\gamma$ is the infimum of the $H_\infty$ design.

\emph{Terminal cost}: To guarantee the monotonous decrease of the cost function $V^o_N$ inside the terminal set, the terminal cost $P_i$ should satisfy
\begin{IEEEeqnarray}{C}
(A_i+B_iK_{i})^TP_i(A_i+B_iK_{i})-P_i\leq-Q-K_i^TRK_i
\label{eq:ric}
\end{IEEEeqnarray}
This condition can be solved efficiently using linear matrix inequalities (LMI). Note that $P_1$ can be calculated easily by taking equality in \eqref{eq:ric} and solve a discrete ARE.

\emph{Terminal set}: The common terminal set $X_f = X_{if}\ (i=1,2)$ is the maximal positively invariant set inside regions B and C and computed based on the gain $K_2$ and the system constraints. For an arbitrary set $Z$, define the operator $\Phi(Z)=\{\bar{x}|\ (A_2+B_2K_2)\bar{x}\in Z\}$. Let $X_0$ be the largest possible compact polyhedron such that 
\begin{IEEEeqnarray}{rCl}
X_0& \subset & \{\bar{x}|\ (x,u)\in \mathbb{X}\times\mathbb{U}\} \cap (\Omega_2\cup\Omega_3).\nonumber\\
X_k& = & \Phi(X_{k-1}) \cap X_{k-1}, \quad i=1,2,...
\label{eq:terset}
\end{IEEEeqnarray}
As proved in \cite{Laz06Stabilizing}, this iterative procedure can be completed in finite step and $X_f=\lim_{k\to\infty}\ X_k$.

The overall MPC controller design is calculated using multiparametric toolbox (MPT) \cite{mpt}.

\section{Simulation Study and Experiment}

\begin{figure}[t]
\centering
\subfloat[]{\includegraphics[width=\columnwidth]{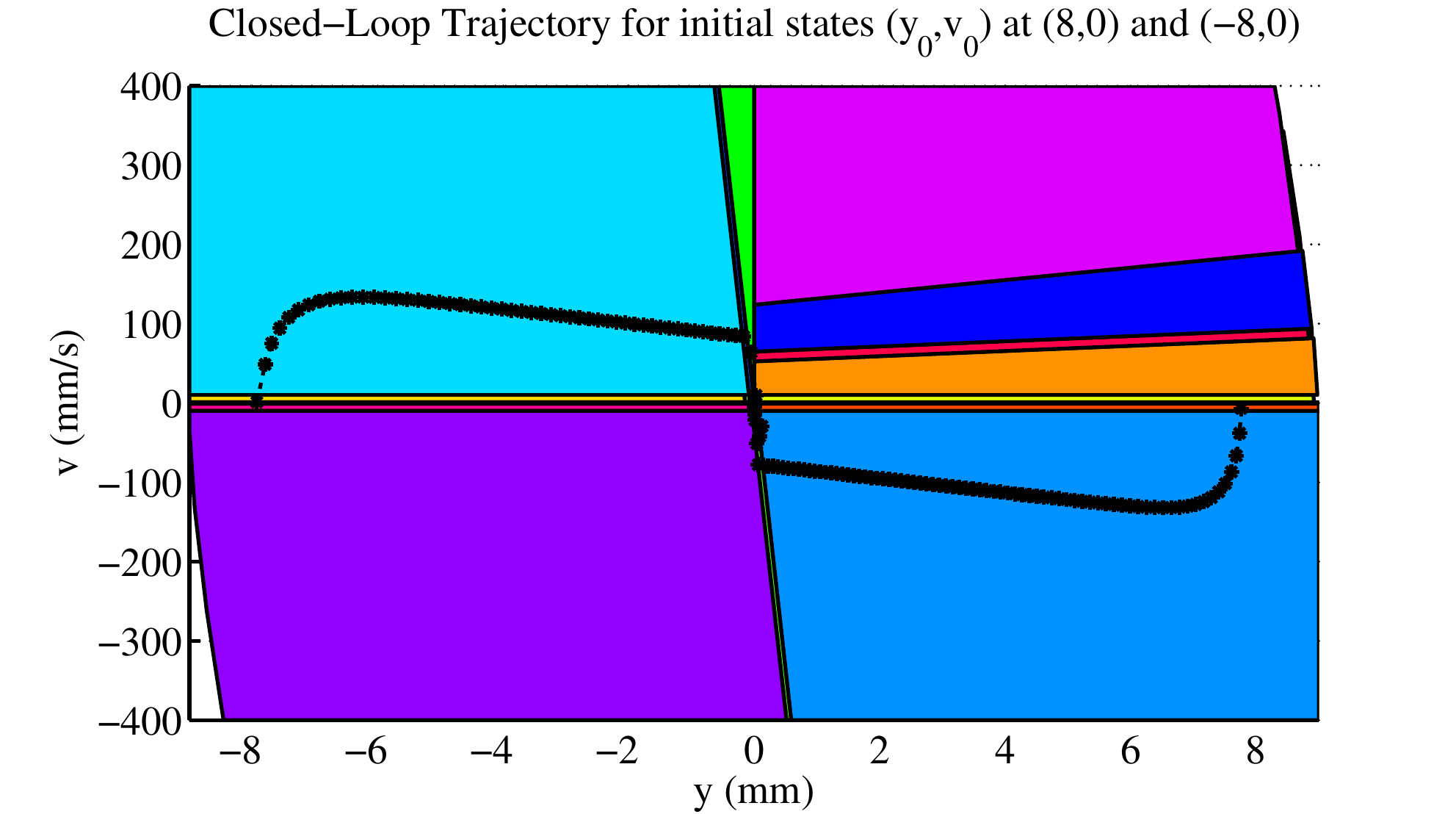}}\\
\subfloat[]{\includegraphics[width=\columnwidth]{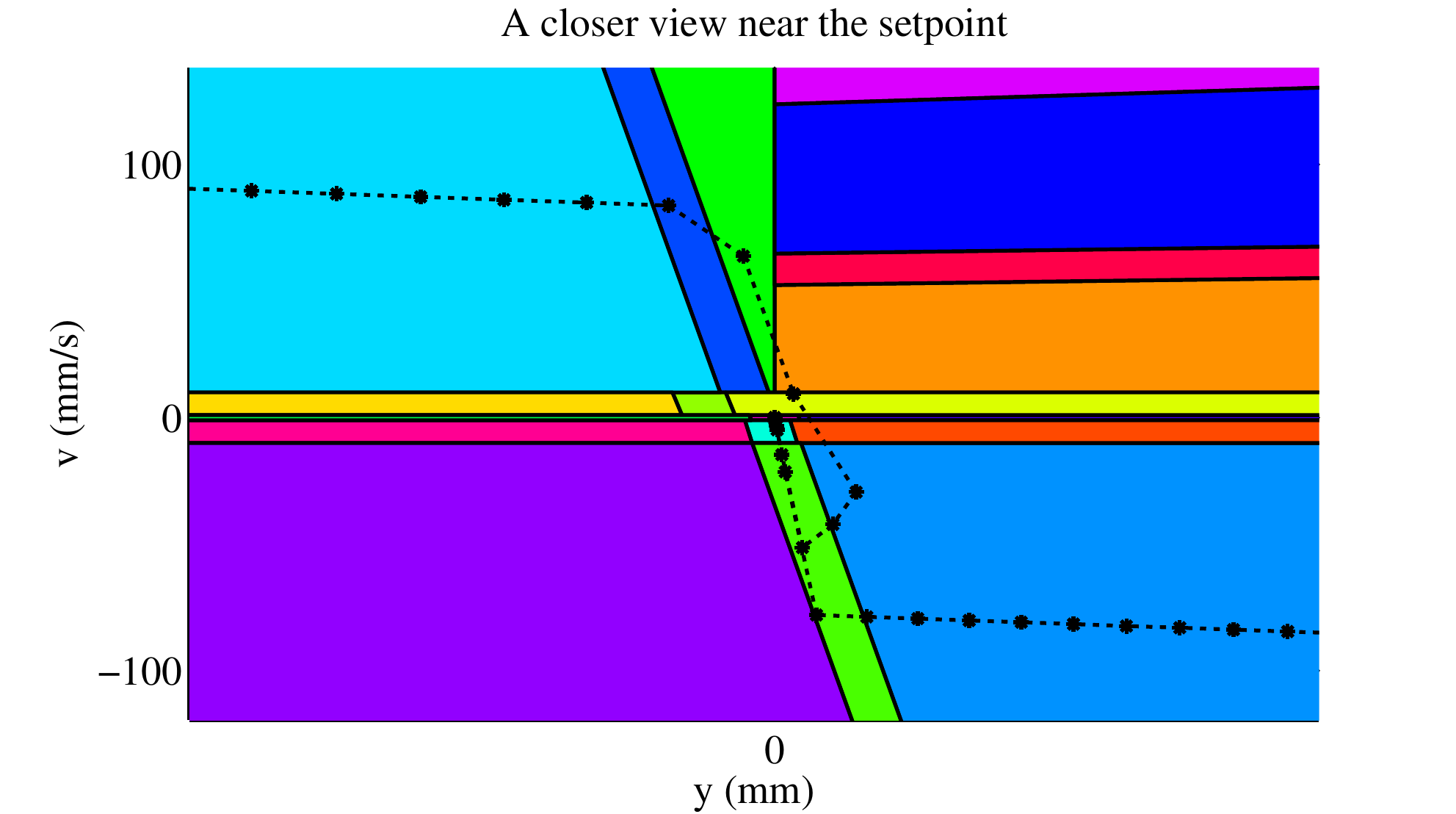}}
\caption{Performance of a normal MPC controller tuned by \cite{Ngu11Enhanced} on the outer model $\{A_1, B_1\}$.}
\label{fig3a}
\end{figure}

\begin{figure}[t]
\centering
\subfloat[]{\includegraphics[width=3.2in]{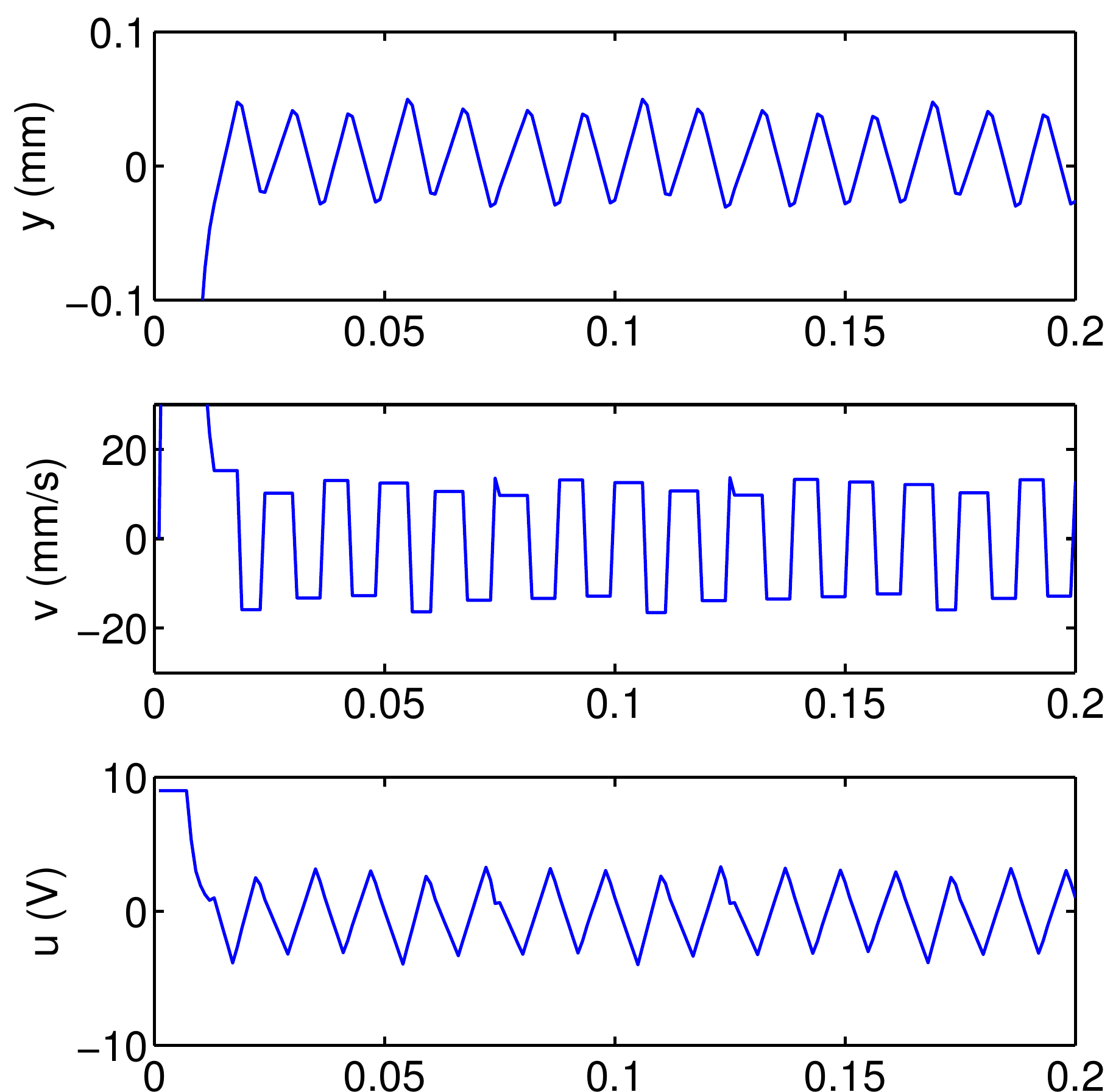}}\\
\subfloat[]{\includegraphics[width=3.2in]{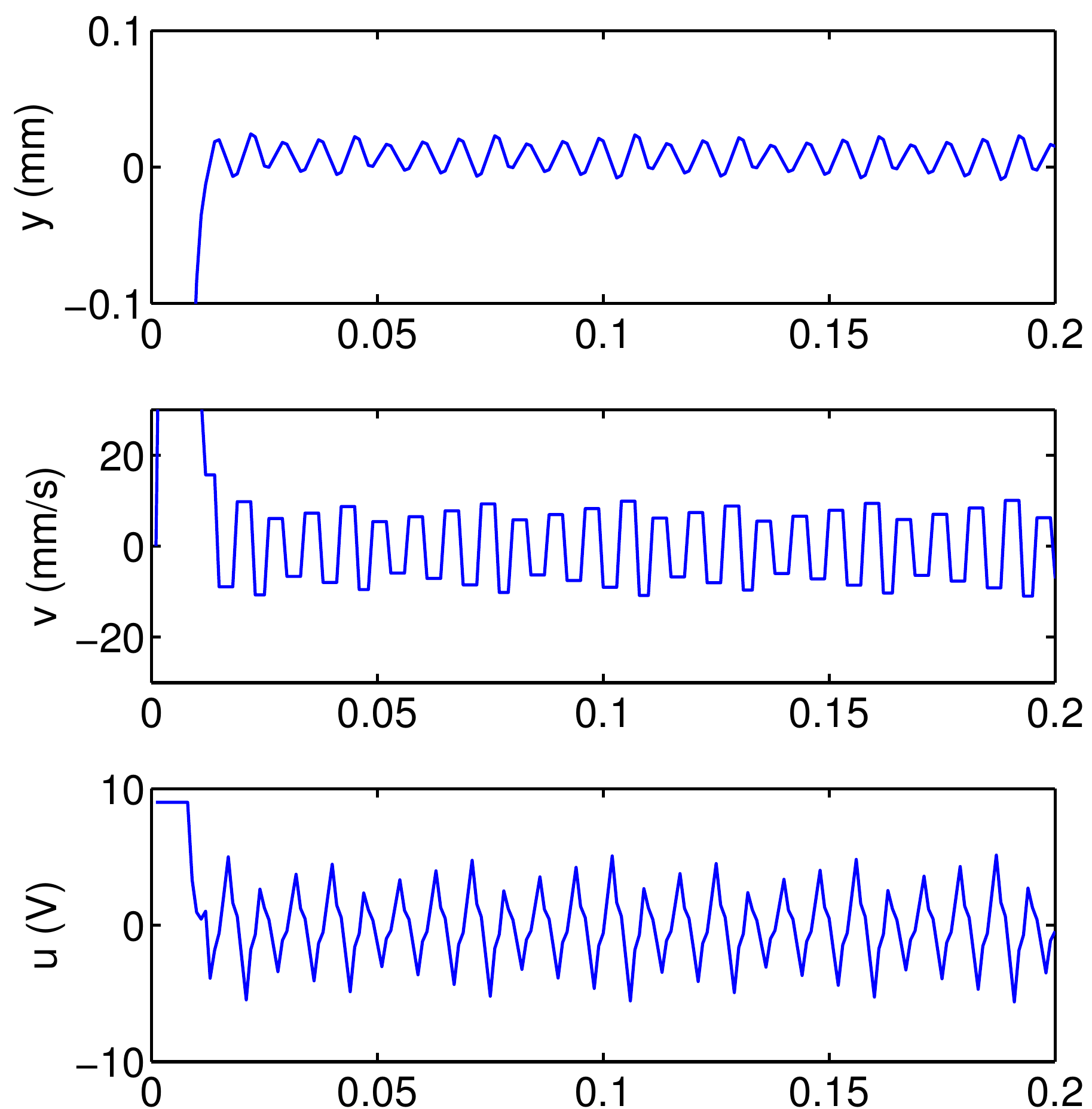}}
\caption{Simulated output errors with friction model mismatch for (a) LQR MPC design and (b) proposed robust MPC.}
\label{fig3}
\end{figure}

\subsection{Simulation Studies}
In this section, the preceding theoretical development is applied to a simulation example. For this study, we assume that the ultrasonic motor has a similar linear model as the models identified from experiment data in Section II, but a different friction form. This choice gives a relative good performance for large changes of set point (Fig. \ref{fig3a}). In fact, the response is fast since $u(k)=u_{max}=10\ V$ only powers up the actuator to a velocity $v(k)=180\ mm/s$, smaller than $v_{max}=400\ mm/s$.

In another study, the robust effective of the prosed control strategy is tested. To simulate a friction uncertainty, we assume that the ultrasonic motor has a similar linear model as the models identified from experiment data in Section II, but a different friction form. The identified parameters are $f_{cp}=2.0V,\ f_{cn}=-2.6V$ and
\begin{IEEEeqnarray}{rCl}
A_1&=&\begin{bmatrix}0.9968& 6.289\times10^{-4}\\ -5.544& 0.3623\end{bmatrix},\ B_1=\begin{bmatrix}4.616\times 10^{-3}\\ 3.493\end{bmatrix}\nonumber\\
A_2&=&\begin{bmatrix}0.9990& 6.312\times 10^{-4}\\ -1.658& 0.3662\end{bmatrix},\ B_2=\begin{bmatrix}2.033\times 10^{-3}\\ 1.636\end{bmatrix}\IEEEeqnarraynumspace
\label{eq:idnmdl}
\end{IEEEeqnarray}
 while the assumed parameters are $f_{cp}=2.4V,\ f_{cn}=-2.9V$, $A_1,B_1$ unchanged and
\begin{IEEEeqnarray}{rCl}
A_2&=&\begin{bmatrix}0.9990& 6.312\times 10^{-4}\\ -1.658& 0.4000\end{bmatrix},\ B_2=\begin{bmatrix}2.033\times 10^{-3}\\ 1.636\end{bmatrix}\IEEEeqnarraynumspace
\label{eq:asmmdl}
\end{IEEEeqnarray}
An increase in $A_2(2,2)$ represents a steeper negative slope of $f_s$ (Fig. \ref{fig1}).

Two MPC schemes are compared: LQR MPC and the proposed controller with the additional robust design. Both MPCs are designed with control horizon $N=5$ and weighting matrices $Q=diag\{10^4,0.5,10^4\},R=0.001$. The choice of $Q,R$ is tuned by the guideline in \cite{Ngu11Enhanced}. Fig. \ref{fig3} shows the tracking errors, velocities and inputs after a step change in the reference. It can be seen that when friction mismatch presents, the output error shows an oscillation around the setpoint. Through the $H_\infty$ design, the oscillation magnitude is substantially reduced from $0.03$ mm to $0.014$ mm. Hence, the effect of mitigating the friction model error is demonstrated by the proposed method.

\subsection{Experiment}

\begin{figure}[ht]
\centering
\includegraphics[width=3.5in]{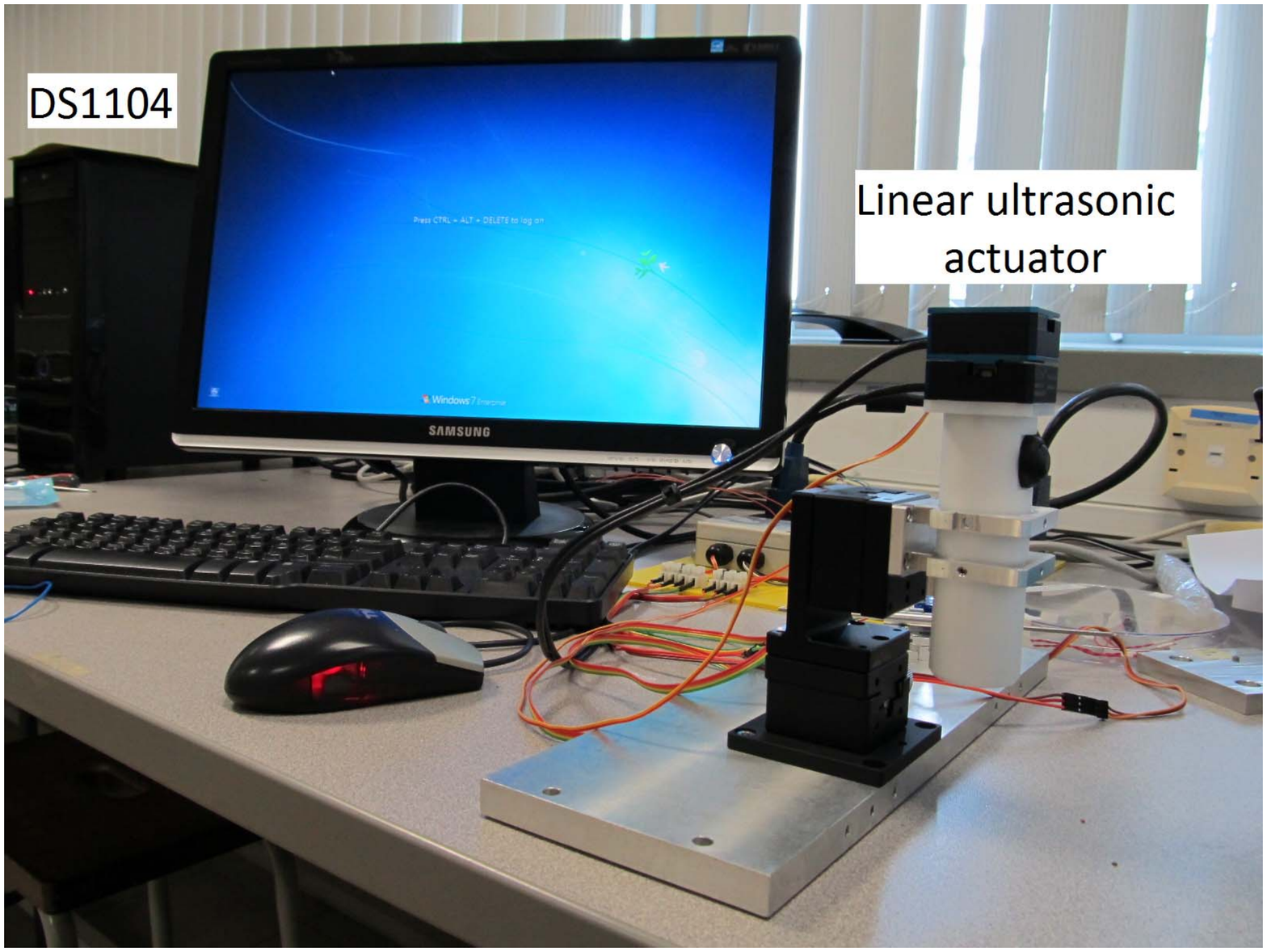}
\caption{Ultrasonic drive control system.}
\label{fig4}
\end{figure}

In this section, real time experiments are also carried out on an ultrasonic drive system (see Fig. \ref{fig4}). The setup uses the PI M-663 with velocity limit $400$ mm/s and travel range $19$ mm. The dSPACE control development and rapid prototyping system, in particular, the DS1104 board, is used, which integrates the whole development cycle seamlessly into a single environment. MATLAB/Simulink can be directly used in dPSACE. The sampling period for this test is chosen as $1$ ms.

The commonly used relay-PID tuning is chosen to compare with the propose method. In this experiment, the system model at $u=3V$ is used for relay tuning to obtain PID gain 
\begin{eqnarray}
K=\begin{bmatrix}34.96&0.09674&1545.5\end{bmatrix}
\label{eq:}
\end{eqnarray}
While it may not offer the best PID tuning in all situation, relay-PID exhibits a large integrating factor to overcome the friction, thus achieving a fast rising time and zero-offset. These characteristics can be used to evaluate the performance of the MPC method.

Parametric programming is used to solve the MPC problem for PWA models. The obtained controller is implemented under a lookup table form with $23$ regions where only matrix multiplication and comparison are performed. 
\begin{eqnarray}
u_k=K_i\bar{x}_k+d_i & \text{if }\bar{x}_k\in CR_i \quad i=1,2,...,23
\label{eq:pwactrl}
\end{eqnarray}
This feasible form is no more complicated than the traditional PID plus nonlinear compensations. Additionally, in order to remove the error oscillation observed in the simulation studies, a dead band with small $\epsilon>0$ is imposed for the input.
\begin{numcases}{u_{k} = }
0 & if $\left|v\right| \leq \epsilon$\nonumber\\
u_{MPC} & if $\left|v\right| > \epsilon$
\label{db}
\end{numcases}
The deadband could be implemented as a pre-condition prior to evaluating \eqref{eq:pwactrl}.

\begin{figure}[t]
\centering
\subfloat[]{\includegraphics[width=3.3in]{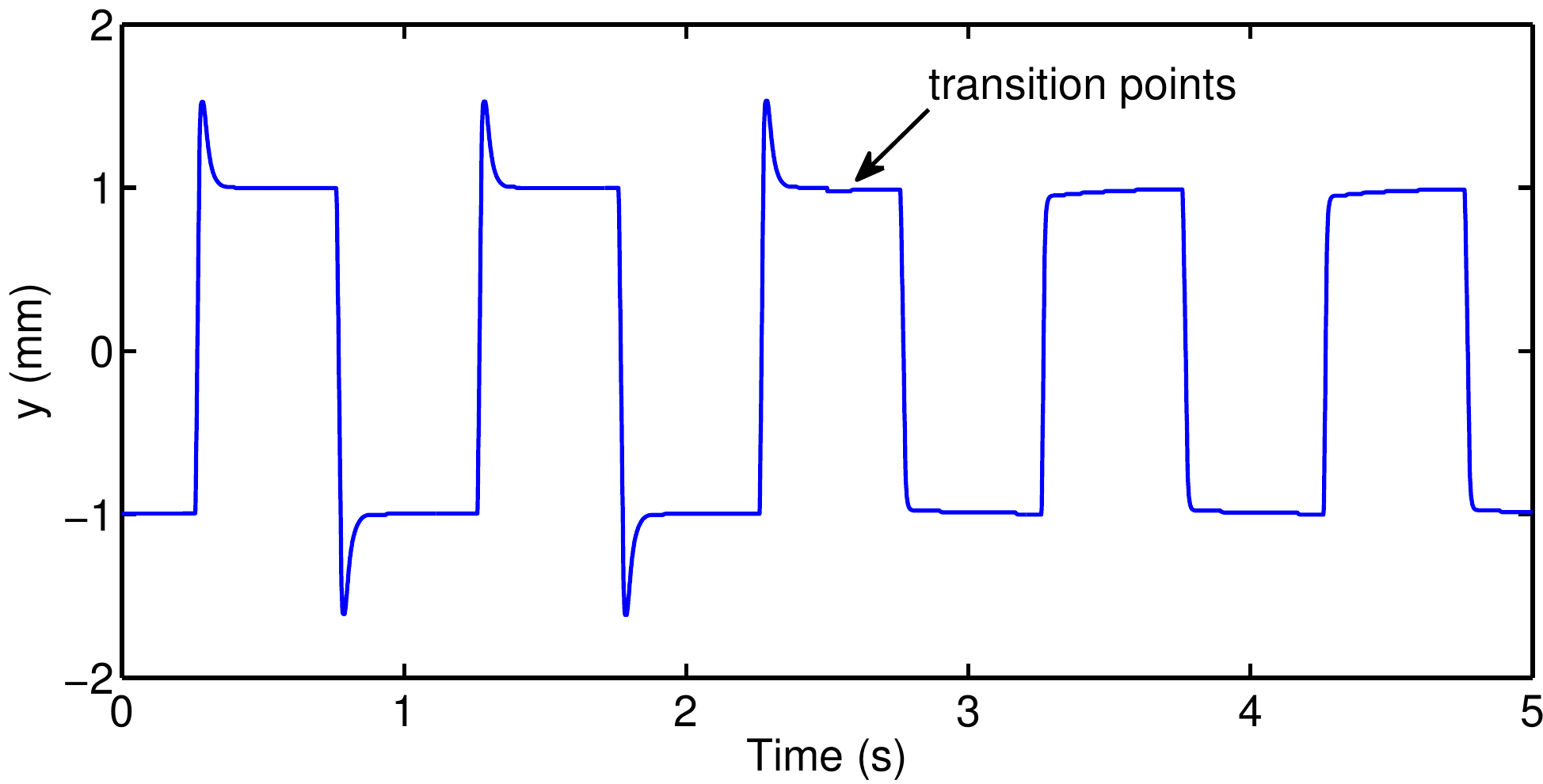}}\\
\subfloat[]{\includegraphics[width=3.4in]{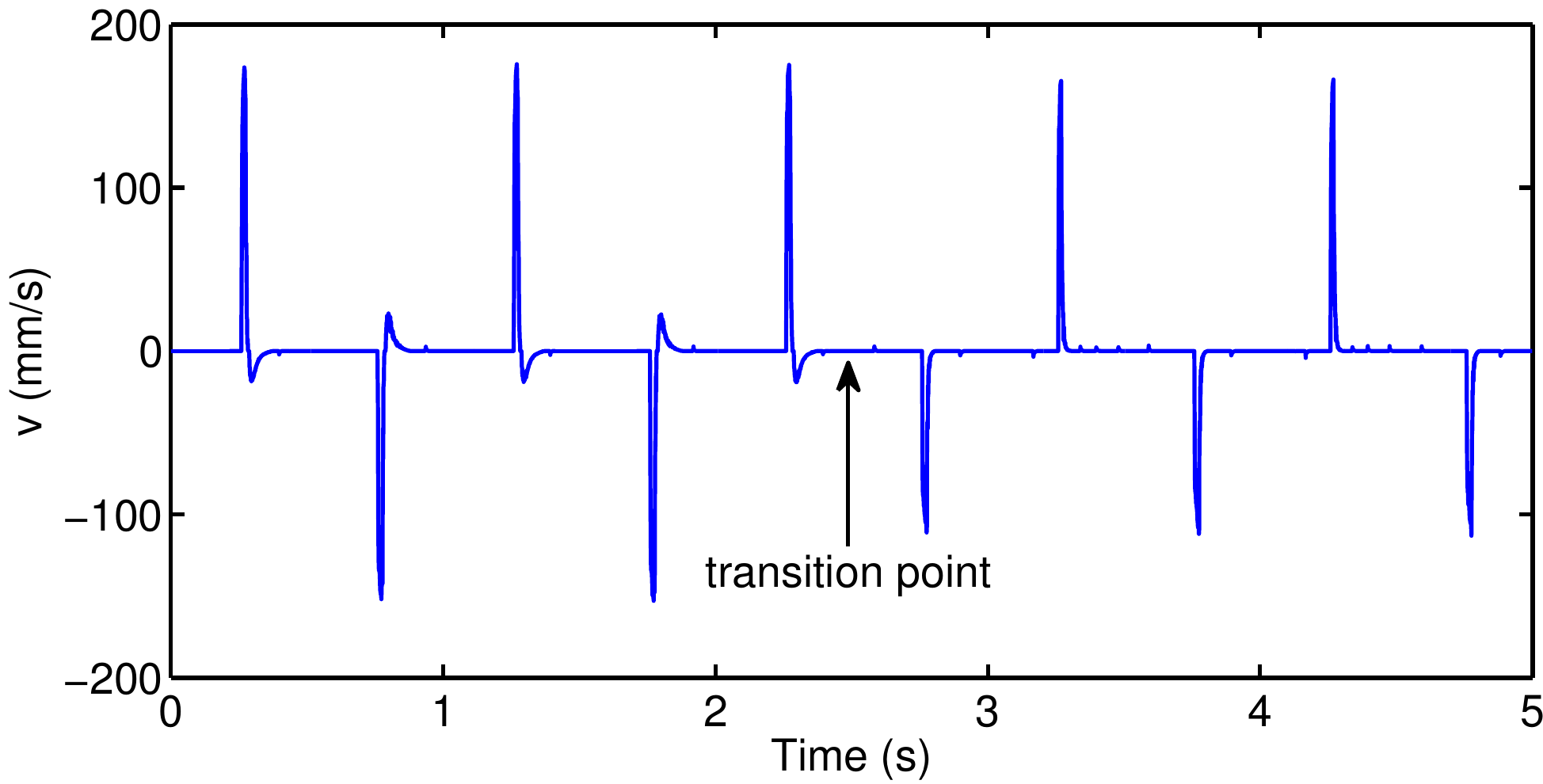}}
\caption{Experiment results when comparing relay-PID (before) and the proposed method (after).}
\label{fig5}
\end{figure}
\addtolength{\textheight}{-6.5cm}
Tracking response of a square wave trajectory ($f=1$ Hz, $A=1$ mm) is shown in Fig. \ref{fig5}. In Fig. \ref{fig5}(a) the tracking output shows that relay-PID creates an overshoot created about $0.5$ mm, which is undesirable. The proposed method can achieve a rising time as fast as the large-integrator PID, and produce no overshoot. Hence, improved steady-state tracking is shown. Fig. \ref{fig5}(b) shows a smooth decreasing of velocity towards zero, implying that the friction uncertainty around the stationary point.

\section{Conclusion}
A robust MPC method has been developed for compensation of friction arising in linear ultrasonic motors. The objective of the control scheme is to achieve good static tracking performance in the presence of the uncertain friction modeling. This is obtained by incorporating linear friction inside the hybrid plant model and designing a robust terminal gain for MPC. Simulation and experimental results have shown that the proposed compensation technique can overcome the limitations of the relay-PID tuning while attaining a simple real-time implementation.

\section*{Acknowledgement}
The authors acknowledge the support from the Collaborative Research Project under the SIMTech-NUS Joint Laboratory (Precision Motion Systems), Ref: U12-R-024JL.

\bibliographystyle{IEEEtran}
\bibliography{IEEEabrv,confRef}

\begin{thebibliography}{10}
\providecommand{\url}[1]{#1}
\csname url@samestyle\endcsname
\providecommand{\newblock}{\relax}
\providecommand{\bibinfo}[2]{#2}
\providecommand{\BIBentrySTDinterwordspacing}{\spaceskip=0pt\relax}
\providecommand{\BIBentryALTinterwordstretchfactor}{4}
\providecommand{\BIBentryALTinterwordspacing}{\spaceskip=\fontdimen2\font plus
\BIBentryALTinterwordstretchfactor\fontdimen3\font minus
  \fontdimen4\font\relax}
\providecommand{\BIBforeignlanguage}[2]{{%
\expandafter\ifx\csname l@#1\endcsname\relax
\typeout{** WARNING: IEEEtran.bst: No hyphenation pattern has been}%
\typeout{** loaded for the language `#1'. Using the pattern for}%
\typeout{** the default language instead.}%
\else
\language=\csname l@#1\endcsname
\fi
#2}}
\providecommand{\BIBdecl}{\relax}
\BIBdecl

\bibitem{Jam09Friction}
Z.~Jamaludin, H.~Van~Brussel, and J.~Swevers, ``Friction compensation of an
  feed table using friction-model-based feedforward and an inverse-model-based
  disturbance observer,'' \emph{Industrial Electronics, IEEE Transactions on},
  vol.~56, no.~10, pp. 3848--3853, 2009.

\bibitem{Par04Identification}
U.~Parlitz, A.~Hornstein, D.~Engster, F.~Al-Bender, V.~Lampaert,
  T.~Tjahjowidodo, S.~D. Fassois, D.~Rizos, C.~X. Wong, K.~Worden, and
  G.~Manson4, ``Identification of pre-sliding friction dynamics,''
  \emph{Chaos}, vol. 14(2), pp. 420--430, 2004.

\bibitem{Pen05Modeling}
K.~Peng, B.~Chen, G.~Cheng, and T.~Lee, ``Modeling and compensation of
  nonlinearities and friction in a micro hard disk drive servo system with
  nonlinear feedback control,'' \emph{Control Systems Technology, IEEE
  Transactions on}, vol.~13, no.~5, pp. 708--721, 2005.

\bibitem{Hay09Discrete}
V.~Hayward, B.~Armstrong, F.~Altpeter, and P.~Dupont, ``Discrete-time
  elasto-plastic friction estimation,'' \emph{Control Systems Technology, IEEE
  Transactions on}, vol.~17, no.~3, pp. 688 --696, may 2009.

\bibitem{Pen07Intelligent}
Y.-F. Peng and C.-M. Lin, ``Intelligent motion control of linear ultrasonic
  motor with h infin; tracking performance,'' \emph{Control Theory
  Applications, IET}, vol.~1, no.~1, pp. 9 --17, january 2007.

\bibitem{Vas07Hybrid}
M.~Vasak, M.~Baotic, I.~Petrovic, and N.~Peric, ``Hybrid theory-based
  time-optimal control of an electronic throttle,'' \emph{Industrial
  Electronics, IEEE Transactions on}, vol.~54, no.~3, pp. 1483 --1494, june
  2007.

\bibitem{Her09Minimum}
M.~Herceg, M.~Kvasnica, and M.~Fikar, ``Minimum-time predictive control of a
  servo engine with deadzone,'' \emph{Control Engineering Practice}, vol.~17,
  no.~11, pp. 1349 -- 1357, 2009.

\bibitem{Laz06Stabilizing}
M.~Lazar, W.~Heemels, S.~Weiland, and A.~Bemporad, ``Stabilizing model
  predictive control of hybrid systems,'' \emph{Automatic Control, IEEE
  Transactions on}, vol.~51, no.~11, pp. 1813 --1818, nov. 2006.

\bibitem{Gad09H-Infinity}
\BIBentryALTinterwordspacing
J.~Gadewadikar, F.~L. Lewis, K.~Subbarao, K.~Peng, and B.~M. Chen, ``H-infinity
  static output-feedback control for rotorcraft,'' \emph{J. Intell. Robotics
  Syst.}, vol.~54, no.~4, pp. 629--646, Apr. 2009. [Online]. Available:
  \url{http://dx.doi.org/10.1007/s10846-008-9279-5}
\BIBentrySTDinterwordspacing

\bibitem{mpt}
M.~Kvasnica, P.~Grieder, and M.~Baoti, ``Multi-parametric toolbox (mpt),''
  2004.

\bibitem{Ngu11Enhanced}
M.~H. Nguyen, K.~K. Tan, and S.~Huang, ``Enhanced predictive ratio control of
  interacting systems,'' \emph{Journal of Process Control}, vol.~21, no.~7, pp.
  1115 -- 1125, 2011.

\end{thebibliography}

\restoregeometry
\end{document}